\documentclass{elsart}

\usepackage{amsmath}
\usepackage{epsf}
\usepackage{graphicx}

\providecommand{\bst}[7]{#1, #2 #3 (#4) #5.} %
\providecommand{\au}[2]{#1 #2}

\providecommand{\IJMPA}{Int. J. Mod. Phys. A} %
\providecommand{\JPG}{J. Phys. G} %
\providecommand{\NPA}{Nucl. Phys. A} %
\providecommand{\NPB}{Nucl. Phys. B} %
\providecommand{\NPBps}{Nucl. Phys. B (Proc. Suppl.)} %
\providecommand{\PLB}{Phys. Lett. B} %
\providecommand{\PPNP}{Prog. Part. Nucl. Phys.} %
\providecommand{\PR}{Phys. Rep.} %
\providecommand{\PRD}{Phys. Rev. D} %
\providecommand{\PRL}{Phys. Rev. Lett.} %

\journal{Physics Letters B, in press}

\begin{document}

\begin{frontmatter}

\title{Sea quark contents of octet baryons}

\author[pku]{Lijing Shao}, \author[lntu]{Yong-Jun
Zhang\corauthref{cor}}\ead{yong.j.zhang@googlemail.com},
\author[pku,chep]{Bo-Qiang Ma\corauthref{cor}} \ead{mabq@phy.pku.edu.cn}
\address[pku]{School of Physics and State Key Laboratory of Nuclear Physics and Technology,
\\Peking University, Beijing 100871, China}
\address[lntu]{Science College, Liaoning Technical University, Fuxin, Liaoning 123000, China}
\address[chep]{Center for High Energy Physics, Peking University, Beijing
100871, China}

\corauth[cor]{Corresponding authors.}


\begin{abstract}
The flavor asymmetry of the nucleon sea, i.e., the excess of
$d\bar{d}$ quark-antiquark pairs over $u\bar{u}$ ones in the proton,
can be explained by several different models; therefore, it is a
challenge to discriminate these models from each other. We examine
in this Letter three models: the balance model, the meson cloud
model, and the chiral quark model, and we show that these models
give quite different predictions on the sea quark contents of other
octet baryons. New experiments aimed at measuring the flavor
contents of other octet baryons are needed for a more profound
understanding of the non-perturbative properties of quantum
chromodynamics (QCD).
\end{abstract}


\end{frontmatter}

\clearpage

\section{Introduction}

The composition of hadrons is one of the central issues of hadronic
physics and can be handled in two languages, i.e., in terms of
quark-gluon degrees of freedom and/or meson-baryon degrees of
freedom. Practically, hadron structures are found to be nontrivial
and more complicated than naive expectations from constitute quark
models. The complications are mainly due to the important
contributions from the non-perturbative behaviors of quantum
chromodynamics (QCD). For instance, as quark-antiquark pairs are created
perturbatively, the sea quarks generated by leading twist evolution,
i.e., from gluon splitting, are necessarily flavor-symmetric and
CP-invariant. Nevertheless, various experiments discovered a
notable flavor asymmetry between $u\bar{u}$ and $d\bar{d}$
quark-antiquark pairs of the
proton~\cite{aea91,aea94,bea94,hea98,pea98,tea01,aea98}. This
flavor asymmetry of the nucleon sea is attributed to the
non-perturbative properties of QCD, and currently, it is still a big challenge
to perform calculations from the first principle of QCD.

 From experimental aspects, the $u\bar{u}$ and $d\bar{d}$ asymmetry was
observed from the violation of the Gottfried sum rule, {\it i.e.},
$S_G = \int_0^1 [(F_2^p - F_2^n)/x] {\rm d}x=1/3$~\cite{g67}, where
$F_2^p$ and $F_2^n$ are the structure functions of the proton and
neutron, respectively, and $x$ is the Bjorken variable, which
measures the fraction of momentum carried by the parton compared to
that of the hadron in the infinite momentum frame (or on the
light-cone). In 1991, the New Muon Collaboration (NMC) utilized
muon-induced deep-inelastic scattering (DIS), and found that $S_G =
0.240 \pm 0.016$~\cite{aea91} (re-evaluated as $0.235 \pm
0.026$~\cite{aea94}). The result is significantly below the
prediction of 1/3 from naive constitute quark considerations. This
was attributed to the flavor asymmetry between $\bar{d}$ and
$\bar{u}$ sea quarks~\cite{prs91}. While the DIS process detects the
difference between $\bar{d}$ and $\bar{u}$ quarks, the Drell-Yan
process can measure their ratios~\cite{es91,r07}. Later, the
deviation was confirmed by the NA51 Collaboration ($\bar{u} /
\bar{d} = 0.51 \pm 0.04 \pm 0.05$ at $x = 0.18$~\cite{bea94}) from
muon pair production through the Drell-Yan process in $p+p$ and
$p+d$ reactions. More accurate ratios of $x$-dependent $\bar{u} /
\bar{d}$ were obtained by the Fermilab E866/NuSea
Collaboration~\cite{hea98,pea98,tea01}, using the 800~GeV protons
interacting with liquid hydrogen and deuterium targets. The HERMES
Collaboration at DESY used an independent method, through
semi-inclusive DIS~\cite{aea98}, and obtained results consistent
with that of the NMC, NA51, and E866/NuSea experiments. Thus, the
flavor asymmetry of light quarks is well established (for reviews,
see Refs.~\cite{k98,gp01}).

 From theoretical aspects, the violation of the Gottfried sum
rule could be alternatively accounted for, at least partially, by
isospin symmetry breaking between the proton and the neutron at the
parton level~\cite{m92,msg93}. To disentangle two possible
scenarios, W$^\pm$ creation, which is free of the $p$-$n$ isospin
symmetry breaking effect, is suggested~\cite{dea94,bs94,pj95}.
Moreover, it has the opportunity to extract $\bar{d} / \bar{u}$
ratios at large $x$ and very high $Q^2$, through the measurements of
ratios of $W^+$, $W^-$ production cross-sections in $p+p$ collisions
at RHIC and LHC~\cite{ypg09}. In this Letter, we assume that the
isospin symmetry between the proton and the neutron holds.

To describe partons inside hadrons, the light-cone presentation is a
natural language~\cite{bpp98}. In the light-cone Fock-state language
of bound states~\cite{bs90,hr97,b00}, the hadronic eigenstate of QCD
Hamiltonian is expanded on the complete set of color-singlet
quark-gluon eigenstates,
\begin{equation}\label{fock}
|h\rangle = \sum_{i,j,k,l} c_{i,j,k,l} | \{q\}, \{i,j,k\}, \{l\}
\rangle \, ,
\end{equation}
where $\{q\}$ represents the valence quarks of the hadron
$|h\rangle$; $i$ is the number of quark-antiquark $u\bar{u}$ pairs;
$j$ is the number of $d\bar{d}$ pairs; $k$ is the
number of gluons; and \{$l$\} denotes other heavier flavors ($s$, $c$,
$b$, and $t$).

It is worthy to mention that the quarks and gluons in the Fock
states are the ``intrinsic'' partons of hadrons, since they are
non-perturbatively multi-connected to valence
quarks~\cite{bea80,bm96}. They are different from the ``extrinsic''
partons generated from QCD hard bremsstrahlung or gluon
splitting as part of the lepton scattering interaction. The
``extrinsic'' sea quarks and gluons only exist for a short time,
$\sim 1/Q$; in contrast, the ``intrinsic'' sea quarks and gluons
exist over a relatively long lifetime within hadronic bound
states. Partons measured at certain $Q^2$ include
both ``intrinsic'' and ``extrinsic'' contents. Since ``extrinsic''
partons are generated without association with flavor structure, the
light flavor sea quark asymmetry mainly originates from
``intrinsic'' partons and is practically $Q^2$-independent or
slightly $Q^2$-dependent~\cite{k98,gp01}.

The initial distributions of nucleon sea flavors are not required
to be symmetric because the nucleon state itself is not CP-invariant.
It is crucial to understand the role of the ``intrinsic'' parton
distributions of hadrons, since they set the boundary conditions for
QCD evolution. Theoretically, there are many phenomenological models
that can account for the flavor asymmetry of the nucleon sea, e.g.,
meson cloud
models~\cite{s72,t83,st87,k91,kfs96,mst98,nea99,mc00,p07,ahm00,hxm04},
chiral quark models~\cite{mg84,w79,ehq92,cl95,br02,dxm05,dxm05b,dm06}, and statistical
models~\cite{zzm01,zzy02,zdm02,zmy03,zsm09}. Besides the $u$ and $d$
flavors, the strange flavor of the nucleon sea has been also
extensively studied~\cite{bm96,dxm05,dxm05b,dm06,ek88,bek88,bw92,n93,dm04}.

While different models give fairly good descriptions of the current
data, measurable differences exist among their predictions,
especially when other members of octet baryons other than nucleons
are considered~\cite{m00}. The quark flavor and spin distributions,
as well as the probabilities to probe them experimentally, are
discussed for $\Lambda^0$~\cite{msy00a,msy00b} and
$\Sigma^\pm$~\cite{aea96,afh98,Ma:1999hi,cs00}. Since $\Lambda^0$ is
charge-neutral and its lifetime is short, it is hard to accelerate
it as an incident beam or use it as a target. Fortunately, various
$\Lambda^0$ fragmentation processes can be used to uncover quark
distributions~\cite{m00,msy00a,msy00b}. As for $\Sigma$, Drell-Yan
experiments with $\Sigma$ beam on protons and deuterium can be
carried out to detect quark
distributions~\cite{m00,aea96,afh98,cs00}.

To examine model-dependent predictions explicitly, in this Letter,
we calculate the sea contents of octet baryons in each of the
different frameworks of the balance model, meson cloud model, and
chiral quark model. We present the different predictions of these
models numerically, for convenience, when comparing the experiments.
We expect new experiments to discriminate the models from each other
and to provide a deeper and more profound understanding of the
flavor structure of hadrons as well as the non-perturbative
behaviors of QCD.

\section{The balance model}


The detailed balance model~\cite{zzm01,zzy02,zmy03} and the balance
model~\cite{zdm02}, which are free from any parameters, were proposed
to look into the statistical effects of the nucleon and to search for
the origin of $d\bar{d}$ and $u\bar{u}$ asymmetry. It was found that
the detailed balance model generates $\bar{d} - \bar{u} \simeq
0.124$, while the balance model gives $\bar{d} - \bar{u} \simeq
0.133$. It is a big surprise that both models provide a remarkable
agreement of their predictions of $d\bar{d}$ over $u\bar{u}$ with
the E866/NuSea result of $0.118 \pm 0.012$~\cite{hea98,pea98,tea01},
without any parameters. Assuming equal probability for every energy
configuration of each $n$-parton Fock state, one can get
$x$-dependent parton distribution functions as well~\cite{zzy02}.
The method was also extended to pions~\cite{ah05} and the nucleon
spin structure~\cite{su04}.

The main idea of the balance model is rather simple and intuitive.
It takes the proton as a bag of quark-gluon gas in dynamical
balance, where partons keep combining and splitting through
processes such as $q(\bar{q}) \Leftrightarrow q(\bar{q})g$, $g
\Leftrightarrow q\bar{q}$, and $g \Leftrightarrow gg$. The model
starts from the valance quark structure of the proton without any
parameters, even the QCD color coupling constant of $\alpha_s$. In this
picture, while $d\bar{d}$ and $u\bar{u}$ sea quark-antiquark pairs
are produced by gluon splitting with equal probability, the reverse
process, i.e., the annihilation of antiquarks with their quark
partners into gluons, is not flavor symmetric due to the net excess
of $u$ quarks over $d$ quarks in the proton. As a consequence,
$\bar{u}$ quarks have a larger probability to annihilate than
$\bar{d}$ quarks, hence bringing an excess of $\bar{d}$ over
$\bar{u}$.

From Eq.~(\ref{fock}), it is easy to see that the probability of
finding a hadron in the Fock state $| \{q\}, \{i,j,k\}, \{l\}
\rangle$ equals to
$\rho_{i,j,k,l} = |c_{i,j,k,l}|^2$,
which satisfies the normalization condition
\begin{equation}\label{norm}
\sum_{i,j,k,l} \rho_{i,j,k,l} = 1 \,.
\end{equation}

\begin{table}
\begin{center}
\caption{Sea contents of octet baryons in the balance
model.}\label{bt}
\begin{tabular}{cccccc}
\hline\hline
Valance quark & Hadron & $\bar{u}$  & $\bar{d}$ & $\bar{d} - \bar{u}$ & $g$ \\
\hline
$uud$   & $p$                       & 0.337 & 0.470 & 0.133 & 1.099\\
$uds$   & $\Lambda^0$               & 0.469 & 0.469 & 0.000 & 1.095\\
$uds$   & $\Sigma^0$                & 0.469 & 0.469 & 0.000 & 1.095\\
$uus$   & $\Sigma^+$                & 0.334 & 0.744 & 0.410 & 1.090\\
$uss$   & $\Xi^0$                   & 0.466 & 0.742 & 0.276 & 1.087\\
\hline
\end{tabular}
\end{center}
\end{table}

In principle, we reasonably assume that the basic property of the
ensemble of the proton does not change with time. As for the detailed
balance model, it is presumed that any two nearby quark-gluon Fock
states should be balanced with each other~\cite{zzm01,zmy03}.
We ignore heavier quarks at first.
The channels from $|\{uud\},\{i,j,k\}\rangle$ to
$|\{uud\},\{i,j,k-1\}\rangle$ are
$ug \xrightarrow{(2+i) \times k} u$, $\bar{u}g \xrightarrow{i \times k} \bar{u}$,
$dg \xrightarrow{(1+j) \times k} d$, $\bar{d}g \xrightarrow{j \times k} \bar{d}$,
and $gg \xrightarrow{C_k^2} g$;
thus,
\begin{equation}
|\{uud\},\{i,j,k\}\rangle
\xrightarrow{(3+2i+2j) \times k+C_k^2}
|\{uud\},\{i,j,k-1\}\rangle \,,
\end{equation}
where $C_k^2 = k(k-1)/2$ and
the number above the arrow is the possible number of the channel.
Inversely, we have
\begin{equation}
|\{uud\},\{i,j,k\}\rangle
\xleftarrow{3+2i+2j+k-1}
|\{uud\},\{i,j,k-1\}\rangle \,.
\end{equation}
The detailed balance condition requires
\begin{equation}\label{gluon}
\rho_{i,j,k} \times [(3+2i+2j) \times k+C_k^2] = \rho_{i,j,k-1} \times (3+2i+2j+k-1) \,.
\end{equation}
Similarly,
\begin{equation}\label{nogluon}
\rho_{i,j,0} \times [i \times (i+2)] = \rho_{i-1,j,1}, \quad
\rho_{i,j,0} \times [j \times (j+1)] = \rho_{i,j-1,1} \,.
\end{equation}
Eqs.~(\ref{norm},~\ref{gluon},~\ref{nogluon}) provide a complete set to solve
the ``intrinsic'' structure of the proton~\cite{zzm01}.

However, there exist some inconsistent points, which are attacked when
a more general principle, named the balance principle, is
adopted~\cite{zdm02}. The balance principle demands that each Fock
state should be balanced with all of its nearby Fock states, not only one Fock state. It induces
a set of linear equations. After including Eq.~(\ref{norm}), we can
determine the parton contents of the proton uniquely as well. For more
details, see Ref.~\cite{zdm02}, which also introduced a method to
include heavier quarks.

The procedure used for the proton is also workable for other
hadrons, and the results for all members of octet baryons, derived
from the balance principle, are listed in Table~\ref{bt}. The
$u$-$d$ isospin symmetry among octet baryons is preserved in our
balance model. Thus, for the neutron $n$ and hyperons $\Sigma^-$ and
$\Xi^-$ (not listed in the table), we can immediately obtain their
parton contents through $u$-$d$ isospin symmetry, e.g., $\bar{u}^n =
\bar{d}^p$, $u^{\Sigma^-} = d^{\Sigma^+}$.

\section{The meson cloud model}

Sullivan~\cite{s72} displayed that virtual meson-baryon states
directly contribute to the nucleon structure. Later,
Thomas~\cite{t83} demonstrated the relevance of the pion cloud for
sea quark distributions, treating SU(3) symmetry as breaking in the
nucleon sea. Further, several authors included $\omega$
meson~\cite{ahm00}, $\sigma$ meson~\cite{hxm04}, as well as pions
and kaons~\cite{s72,t83,st87,k91,kfs96,mst98,nea99,mc00,p07}. We now
refer to these models as meson cloud models. However, in our
calculation, only pions, which contribute to structure functions
most significantly due to their lightest mass, are considered. For
the same reason, only baryons in the octet and decuplet states are
taken into account in this Letter.

The proton has virtual states such as $\pi N$ and $\pi \Delta$. Here
we write its wavefunction as follows~\cite{gp01},
\begin{eqnarray}
|p\rangle \rightarrow \sqrt{1-a-b}~|p_0\rangle & + & \sqrt{a}~\left(
-\sqrt{\frac{1}{3}} |p_0\pi^0\rangle + \sqrt{\frac{2}{3}}
|n_0\pi^+\rangle \right) \nonumber \\
 &+& \sqrt{b}~\left( \sqrt{\frac{1}{2}}
|\Delta^{++}_0 \pi^-\rangle - \sqrt{\frac{1}{3}} |\Delta^+_0
\pi^0\rangle \nonumber + \sqrt{\frac{1}{6}} |\Delta^0_0 \pi^+\rangle
\right) \,,
\end{eqnarray}
where the subscript ``0'' denotes the bare part, or equivalently speaking,
where only valence quarks are involved; the coefficients inside the
brackets are from isospin couplings~\cite{aea08}; $a$ and $b$ are
weight factors for states from the baryon octet and decuplet states,
respectively, satisfying
\begin{equation}\label{abab}
a > 0, \quad b > 0, \quad a + b < 1 \,.
\end{equation}

We impose another constraint,
\begin{equation}\label{agb}
a > b \,,
\end{equation}
considering the fact that baryons in the decuplet state are heavier than
those in the octet state; thus, they are suppressed.

From the wavefunction of the proton, we get the $\bar{d}$ and
$\bar{u}$ contents directly,
\begin{equation}\label{pd}
\bar{d} = \frac{5}{6} a + \frac{1}{3} b \,,
\end{equation}
\begin{equation}\label{pu}
\bar{u} = \frac{1}{6} a + \frac{2}{3} b \,.
\end{equation}

\begin{table}
\begin{center}
\caption{Expressions of sea contents of octet baryons in the meson
cloud model. The numbers inside the brackets are typical values when a
commonly used relation, $a=2b$~\cite{gp01}, is adopted.}\label{ptex}
\begin{tabular}{ccccccc}
\hline\hline
Hadron      & \multicolumn{2}{c}{$\bar{u}$}&\multicolumn{2}{c}{$\bar{d}$}&\multicolumn{2}{c}{$\bar{d}-\bar{u}$}\\
\hline
$p$         & $1/6a + 2/3b$&$(0.130)$ & $5/6a + 1/3b$&$(0.260)$ & $2/3a - 1/3b$&$(0.130)$\\
$\Lambda^0$ & $1/2a + 1/2b$&$(0.195)$ & $1/2a + 1/2b$&$(0.195)$ & $0$&$(0.000)$          \\
$\Sigma^0$  & $a + 1/2b$&$(0.325)$    & $a + 1/2b$&$(0.325)$    & $0$&$(0.000)$          \\
$\Sigma^+$  & $1/4a + 1/4b$&$(0.098)$ & $7/4a + 3/4b$&$(0.553)$ & $3/2a + 1/2b$&$(0.455)$\\
$\Xi^0$     & $1/6a + 1/6b$&$(0.065)$ & $5/6a + 5/6b$&$(0.325)$ & $2/3a + 2/3b$&$(0.260)$\\
\hline
\end{tabular}
\end{center}
\end{table}

\begin{figure}
\begin{center}
\includegraphics[width=10cm]{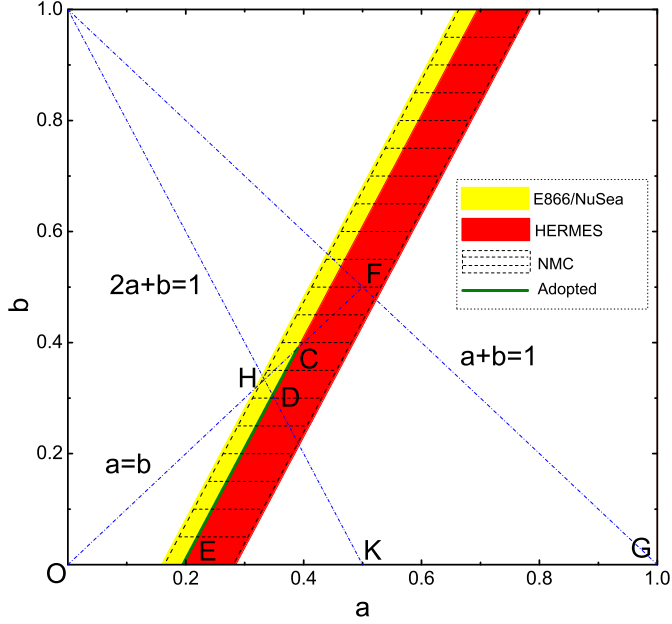}\caption{Determinations of parameters
$a$ and $b$ in the meson cloud model. The dash-dotted lines limit
$a$ and $b$ to the bottom triangle
according to Eqs.~(\ref{abab},\ref{agb},\ref{abab2}):
$\triangle$OFG for nucleons,
$\Lambda$, $\Xi$, and $\triangle$OHK for $\Sigma$. The shaded area,
light colored area, and dark colored area are results
from NMC~\cite{aea91,aea94},
E866/NuSea~\cite{hea98,pea98,tea01}, and HERMES~\cite{aea98},
respectively. The heavy line represents the scope of parameters we
ultimately adopt in our calculations: the segment $\overline{\mathrm{CE}}$
for nucleons, $\Lambda$, $\Xi$, and the segment $\overline{\mathrm{DE}}$ for
$\Sigma$.\label{ab}}
\end{center}
\end{figure}

As for $\bar{d}$-$\bar{u}$ asymmetry, the NMC experiment gave $\bar{d} -
\bar{u} = 0.148 \pm 0.039$~\cite{aea91,aea94}, while the E866/NuSea
Collaboration reported $\bar{d} - \bar{u} = 0.118 \pm
0.012$~\cite{hea98,pea98,tea01} and the HERMES Collaboration obtained
$\bar{d} - \bar{u} = 0.16 \pm 0.03$~\cite{aea98}. They are
illustrated in Fig.~\ref{ab} in the shaded area,
light colored area, and dark colored area, respectively.
For convenience, we adopt
$\bar{d} - \bar{u} = 0.130$ in our following calculations,; thus, the
three experimental results are all satisfied with errors considered.
Thereafter, by substituting Eq.~(\ref{pd}) and Eq.~(\ref{pu}), we
reach a relation of $a$ and $b$, as
\begin{equation}\label{dmu}
\bar{d} - \bar{u} = \frac{1}{3} (2a-b) = 0.130 \,,
\end{equation}
which
will later be used extensively as an experimental constraint.

After combining the constraints, i.e.,
Eqs.~(\ref{abab},~\ref{agb},~\ref{dmu}), we can set down the
relation between $a$ and $b$, as well as the boundary, as
\begin{equation}\label{con1}
b = 2a - 0.39, \quad a \in (0.195,0.390)\,,
\end{equation}
which is shown in Fig.~\ref{ab} as the segment $\overline{\mathrm{CE}}$.

Further, within the same framework, we also explicitly write down the
wavefunctions for $\Lambda^0$, $\Sigma^0$, $\Sigma^+$, and $\Xi^0$
,
\begin{eqnarray}
|\Lambda^0\rangle \rightarrow \sqrt{1-a-b}~|\Lambda^0_0\rangle & + &
\sqrt{a}~\left( \sqrt{\frac{1}{3}} |\Sigma^+_0 \pi^- \rangle -
\sqrt{\frac{1}{3}} |\Sigma^0_0 \pi^0 \rangle +
\sqrt{\frac{1}{3}} |\Sigma^-_0 \pi^+ \rangle \right)  \nonumber \\
& + & \sqrt{b}~\left( \sqrt{\frac{1}{3}} |\Sigma^{*+}_0 \pi^-
\rangle -\sqrt{\frac{1}{3}}|\Sigma^{*0}_0 \pi^0 \rangle +
\sqrt{\frac{1}{3}} |\Sigma^{*-}_0\pi^+\rangle \right) \,, \nonumber
\end{eqnarray}
\begin{eqnarray}
|\Sigma^0\rangle \rightarrow \sqrt{1-2a-b}~|\Sigma^0_0\rangle & +
&\sqrt{a}~\left( \sqrt{\frac{1}{2}} |\Sigma^+_0\pi^-\rangle -
\sqrt{\frac{1}{2}} |\Sigma^-_0\pi^+\rangle \right) + \sqrt{a}~|\Lambda^0_0 \pi^0 \rangle \nonumber \\
& + & \sqrt{b}~\left( \sqrt{\frac{1}{2}} |\Sigma^{*+}_0\pi^-\rangle
- \sqrt{\frac{1}{2}}|\Sigma^{*-}_0\pi^+\rangle \right) \,, \nonumber
\end{eqnarray}
\begin{eqnarray}
|\Sigma^+\rangle \rightarrow \sqrt{1-2a-b}~|\Sigma^+_0\rangle & +
&\sqrt{a}~\left( \sqrt{\frac{1}{2}} |\Sigma^+_0\pi^0\rangle -
\sqrt{\frac{1}{2}} |\Sigma^0_0\pi^+\rangle \right) + \sqrt{a}~|\Lambda^0_0 \pi^+ \rangle \nonumber \\
& + &\sqrt{b}~\left( \sqrt{\frac{1}{2}} |\Sigma^{*+}_0\pi^0\rangle -
\sqrt{\frac{1}{2}}|\Sigma^{*0}_0\pi^+\rangle \right) \,, \nonumber
\end{eqnarray}
\begin{eqnarray}
|\Xi^0\rangle \rightarrow \sqrt{1-a-b}~|\Xi^0_0\rangle & +
&\sqrt{a}~\left( -
\sqrt{\frac{1}{3}}|\Xi^0_0\pi^0\rangle +
\sqrt{\frac{2}{3}} |\Xi^-_0\pi^+\rangle  \right) \nonumber \\
& + &\sqrt{b}~\left( - \sqrt{\frac{1}{3}}|\Xi^{*0}_0\pi^0\rangle +
\sqrt{\frac{2}{3}} |\Xi^{*-}_0\pi^+\rangle \right) \,. \nonumber
\end{eqnarray}

Through the same procedure, we derive the $\bar{d}$ and $\bar{u}$ contents
inside the above baryons. The expressions of these results in terms of
$a$ and $b$ are listed in Table~\ref{ptex}.

One should caution that a new constraint,
\begin{equation}\label{abab2}
a > 0, \quad b > 0, \quad 2a + b < 1
\end{equation}
should replace Eq.~(\ref{abab}) for $\Sigma$. Thereafter, the former
constraint, the segment $\overline{\mathrm{CE}}$ in Fig.~\ref{ab}, is
replaced by the segment $\overline{\mathrm{DE}}$ for $\Sigma$. Equivalently,
Eq.~(\ref{con1}) changes into
\begin{equation}\label{con2}
b = 2a - 0.39, \quad a \in (0.195,0.348) \,.
\end{equation}

From Table~\ref{ptex} and the corresponding constraints,
Eq.~(\ref{con1}) for nucleons, $\Lambda$, $\Xi$, and
Eq.~(\ref{con2}) for $\Sigma$, we finally arrive at numerical results
for the $\bar{d}$ and $\bar{u}$ sea quarks, as shown in Table~\ref{ptnu}.

\begin{table}
\begin{center}
\caption{Numerical results of sea contents of octet baryons in the
meson cloud model.}\label{ptnu}
\begin{tabular}{cccc}
\hline\hline
Hadron & $\bar{u}$  & $\bar{d}$ & $\bar{d} - \bar{u}$\\
\hline
$p$                       & $(0.033,0.325)$ & $(0.163,0.455)$ & $0.130$\\
$\Lambda^0$               & $(0.098,0.390)$ & $(0.098,0.390)$ & $0.000$\\
$\Sigma^0$                & $(0.195,0.501)$ & $(0.195,0.501)$ & $0.000$\\
$\Sigma^+$                & $(0.049,0.164)$ & $(0.341,0.839)$ & $(0.293,0.675)$\\
$\Xi^0$                   & $(0.033,0.130)$ & $(0.163,0.650)$ & $(0.130,0.520)$\\
\hline
\end{tabular}
\end{center}
\end{table}

\section{The chiral quark model}

In the chiral quark model, the mesons are emitted by valence
quarks~\cite{mg84,w79,ehq92,cl95,br02,dxm05,dxm05b,dm06} instead of baryons in the
meson cloud model. The $u$ and $d$ flavors of hadrons can be read as
\begin{eqnarray}
|u\rangle \rightarrow \sqrt{1-w}~|u_0\rangle + \sqrt{w}~\left(
-\sqrt{\frac{1}{3}} |u_0 \pi^0\rangle +
\sqrt{\frac{2}{3}}|d_0\pi^+\rangle \right) \,,\nonumber
\end{eqnarray}
\begin{eqnarray}
|d\rangle \rightarrow \sqrt{1-w}~|d_0\rangle + \sqrt{w}~\left(
-\sqrt{\frac{2}{3}} |u_0 \pi^-\rangle +
\sqrt{\frac{1}{3}}|d_0\pi^0\rangle \right) \,,\nonumber
\end{eqnarray}
where $w$ is a weight factor indicating the probability of emitting
pions. At a first approximation, the strange quarks are assumed to
be highly suppressed (i.e., no emissions) because of their heavier
mass. The above expressions are equivalent to
\begin{equation}
u \rightarrow \left( 1 + \frac{w}{6} \right) u + \frac{5w}{6} d +
\frac{w}{6} \bar{u} + \frac{5w}{6} \bar{d} \,,
\end{equation}
\begin{equation}
d \rightarrow \left( 1 + \frac{w}{6} \right) d + \frac{5w}{6} u +
\frac{w}{6} \bar{d} + \frac{5w}{6} \bar{u} \,.
\end{equation}

For the proton $|uud\rangle$, we have
\begin{equation}
\bar{d} - \bar{u} = \frac{11}{6} w - \frac{7}{6} w = \frac{2}{3} w
\,.
\end{equation}
Again, we adopt the experimental constraint of $\bar{d} - \bar{u} =
0.130$, as discussed previously, thereby making $w = 0.195$. The analytical
results together with the numerical results for octet baryons are listed
in Table~\ref{ct}.

\begin{table}
\begin{center}
\caption{Sea contents of octet baryons in the chiral quark
model.}\label{ct}
\begin{tabular}{cccc}
\hline\hline
Hadron & $\bar{u}$  & $\bar{d}$ & $\bar{d} - \bar{u}$\\
\hline
$p$                       & $7/6w = 0.228$ & $11/6w = 0.358$ & $2/3w = 0.130$\\
$\Lambda^0$               & $w = 0.195$ & $w = 0.195$ & $0.000$          \\
$\Sigma^0$                & $w = 0.195$    & $w = 0.195$    & $0.000$          \\
$\Sigma^+$                & $1/3w = 0.065$ & $5/3w = 0.325$ & $4/3w = 0.260$\\
$\Xi^0$                   & $1/6w = 0.033$ & $5/6w = 0.163$ & $2/3w = 0.130$\\
\hline
\end{tabular}
\end{center}
\end{table}

\section{Discussion}

In order to get a physical understanding of hadron structure,
and to explain experimental data,
people have suggested many phenomenological models for hadrons in
terms of quark-gluon degrees of freedom and meson-baryon degrees of
freedom. Here we considered three candidate models in their simple
versions, and accomplished an extensive study of the sea contents
for octet baryons. The results are listed in
Tables~\ref{bt},~\ref{ptex},~\ref{ptnu}, and ~\ref{ct}.

The meson cloud model of simple version only presents wide ranges
for its parameters, due to our simple consideration of obvious
constraints. However, to compare between models more conveniently,
we introduce another commonly used relation between parameters $a$
and $b$~\cite{gp01},
\begin{equation}
a = 2 b \, .
\end{equation}
Using this relation, together with Eq.~(\ref{dmu}), we can
determine $a$ and $b$ straightly, as
\begin{equation}
a = 0.26 \,, \quad b = 0.13 \,.
\end{equation}
The results for this special case are listed in Table~\ref{ptex}
within the brackets. Therefore, the fluctuations of octet baryons into
meson-baryon states contribute $a +b \sim 39\%$ for nucleons,
$\Lambda$, $\Xi$, and $2a+b \sim 65\%$ for $\Sigma$, which are
rather significant.

\begin{figure}
\begin{center}
\includegraphics[width=10cm]{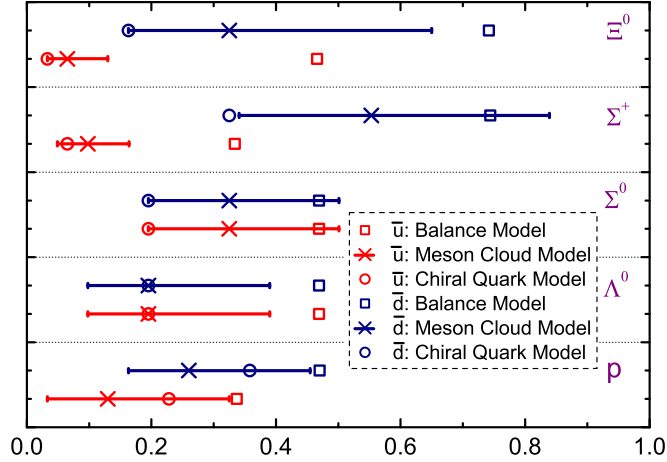}\caption{Sea contents predicted by three
models for five baryons. The horizontal axis is the integrated
number of the $\bar{d}$ and $\bar{u}$ contents.
The dark marks and lines stand for the $\bar{d}$ quarks,
while the light marks and lines represent the $\bar{u}$ quarks. The horizontal
lines stand for the ranges predicted by the meson cloud model.
The squares, crosses, and circles are
predictions for the balance model, a typical case of meson
cloud model with $a=2b$, and the chiral quark model, respectively.
\label{gud}}
\end{center}
\end{figure}

The predicted results of the three models, together with the above
typical case, are illustrated in Fig.~\ref{gud}.
The horizontal axis is the integrated number of
$\bar{d}$ and $\bar{u}$ contents.
The dark marks and lines stand for the $\bar{d}$ quarks,
while the light marks and lines represent the $\bar{u}$ quarks. The horizontal
lines stand for the ranges predicted by the meson cloud model.
The squares, crosses, and circles are
predictions of the balance model, a typical case of meson
cloud model with $a=2b$, and the chiral quark model, respectively.

In Fig.~2, we see many differences among the models. The balance
model, for most of the time, gives the maximum sea contents for the
$\bar{d}$ and $\bar{u}$ quarks, especially for $\bar{d}$ inside the
proton, $\Lambda^0$, $\Xi^0$, and $\bar{u}$ inside the proton,
$\Lambda^0$, $\Sigma^0$, $\Xi^0$. As for $\Sigma^+$ and $\Xi^0$, the
balance model presents a remarkably large $\bar{u}$ sea, compared to
other models. Also worthy of mentioning, the balance model can
predict gluons as well, as shown in Table~\ref{bt}. If we just
consider the $a=2b$ case to stand for the meson cloud model, it is
found that with the number of strange quarks increasing, the sea
contents predicted from the meson cloud model become larger related
to the chiral quark model, albeit always smaller than the balance
model. For $\Lambda^0$ and $\Sigma^0$, three models all give a
symmetric sea, and no $\bar{d}$ and $\bar{u}$ asymmetry is
predicted; this can be used to experimentally test the robustness of
all three models. Very tiny $\bar{u}$ sea quarks are predicted for
$\Sigma^+$ and $\Xi^0$, except by the balance model; thus, this also
provides a window to discriminate models from each other.

\section{Summary}

Flavor structure, in terms of quark-gluon degrees of freedom or
meson-baryon degrees of freedom, is of great interest among the hadronic
society, mainly due to the nontrivial and complicated contents of
sea quarks, e.g., $d\bar{d}$ and $u\bar{u}$ asymmetry, originating
from multi-connected, non-perturbative quantum chromodynamics (QCD).
Because there calculations still remain difficult when the
perturbative assumption falls down, many phenomenological models are
raised to account for the experimental results. However, their
predictive powers appear successful at some places while not so
satisfactory at other places; hence, it is hard to decide which one
is better at describing the flavor content of hadrons.

As suggested, new domains of $\Lambda$ physics and $\Sigma$ physics
could provide plentiful opportunities to discriminate models from
each other and to search for the more profound nature of hadronic
physics. While nucleons have been studied extensively both
experimentally and theoretically, other hadrons still need more
investigation.

In this Letter, we present an extensive study on the sea contents of
octet baryons based on the balance model, meson cloud model, and
chiral quark model. Numerical results are given explicitly, and
these results can be used to distinguish models from each other once
relevant experiments become available. The difference between models
is significant; hence, new experiments aimed at determining the sea
content of other members of octet baryons can open windows to test
different scenarios of the sea content of baryons.

\section*{Acknowledgments}

This work is supported by the National Natural Science Foundation of
China (Nos. 10721063, 10975003). It is also supported by the Hui-Chun
Chin and Tsung-Dao Lee Chinese Undergraduate Research Endowment
(Chun-Tsung Endowment) at Peking University and by the National Fund
for Fostering Talents of Basic Science (Nos. J0630311, J0730316).



\end{document}